\documentclass[10pt,twocolumn,twoside]{IEEEtran}

\usepackage{amsfonts}
\usepackage{amssymb}
\usepackage{amstext}
\usepackage{amsthm}
\usepackage{graphicx}
\usepackage{url}
\usepackage{epstopdf}
\usepackage{color}
\usepackage{xcolor}
\usepackage{bbm}
\usepackage[cmex10]{amsmath}
\usepackage{array}
\usepackage{caption}
\usepackage{cite}
\colorlet{mygreen}{green!60!gray}

\newtheorem{definition}{Definition}

\renewcommand{\SS}{\mathcal S}

\begin{document}
%
\title{A Game-Theoretic Framework for Optimum Decision Fusion in the Presence of Byzantines}
%
%
%

\author{Andrea Abrardo,~ Mauro Barni,~\IEEEmembership{Fellow,~IEEE},~Kassem Kallas,~\IEEEmembership{Student Member,~IEEE}, Benedetta Tondi,~\IEEEmembership{Student Member,~IEEE}
\thanks{A. Abrardo, M. Barni, K. Kallas, B. Tondi are with Dept. of Information Engineering and Mathematical Sciences, Universit\`a di Siena, Italy (barni@dii.unisi.it, abrardo@dii.unisi.it, k\_kallas@hotmail.com, benedettatondi@gmail.com).}
}
\maketitle

\begin{abstract}
Optimum decision fusion in the presence of malicious nodes - often referred to as Byzantines - is hindered by the necessity of exactly knowing the statistical behavior of Byzantines. By focusing on a simple, yet widely studied, set-up in which a Fusion Center  (FC) is asked to make a binary decision about a sequence of system states by relying on the possibly corrupted decisions provided by local nodes, we propose a game-theoretic framework which permits to exploit the superior performance provided by optimum decision fusion, while limiting the amount of a-priori knowledge required. We first derive the optimum decision strategy by assuming that the statistical behavior of the Byzantines is known. Then we relax such an assumption by casting the problem into a game-theoretic framework in which the FC tries to {\em guess} the behavior of the Byzantines, which, in turn, must fix their corruption strategy without knowing the {\em guess} made by the FC. We use numerical simulations to derive the equilibrium of the game, thus identifying the optimum behavior for both the FC and the Byzantines, and to evaluate the achievable performance at the equilibrium. We analyze several different setups, showing that in all cases the proposed solution permits to improve the accuracy of data fusion. We also show that, in some instances, it is preferable for the Byzantines to minimize the mutual information between the status of the observed system and the reports submitted to the FC, rather than always flipping the decision made by the local nodes as it is customarily assumed in previous works.
\end{abstract}


\IEEEpeerreviewmaketitle

\section{Introduction}
\label{sec.intro}
Decision fusion for distributed detection in the presence of malicious nodes, often referred to as Byzantines \cite{Vemp13}, has received an increasing attention for its importance in several application scenarios, including wireless sensor networks \cite{WSNDDByz,WSNanomalyDet}, cognitive radio \cite{WLSH10,Raw11,Zhang2013secure,wang2014secure}, distributed detection \cite{Mar09,DistrDetTree}, multimedia forensics \cite{Bar13} and many others.

The most commonly studied scenario is the parallel distributed data fusion model. According to such a model, the $n$ nodes of a multi-sensor network collect information about a system through the observation vectors ${\bf x}_1, {\bf x}_2 \dots {\bf x}_n$. Based on the observation vectors, the nodes compute $n$ reports and send them to a Fusion Center (FC). The fusion center gathers the local reports and makes a final decision about the state of the system. In the setup considered in this paper, the state of the system is represented by a sequence $s^m = (s_1, s_2 \dots s_m)$. The $m$ components of $s^m$ may correspond to the state of the system at different time instants or to several characteristics of a complex system. Spectrum sensing in cognitive radio networks is a typical example of the above scenario: the observed system is the frequency spectrum of a primary communication network, while $s^m$ may correspond to the state of the spectrum (busy or idle) at different time instants, or to the state of different frequency channels. An additional example is provided by online reputation systems. In such systems, the FC must compute the global rating of a good or a service by relying on the feedback and ratings coming from users, whom, in turn, could be interested to provide a biased feedback in order to increase or decrease the reputation of an item \cite{RepSys}.

Hereafter we assume that each component of $s^m$ can take only two values ($s_i \in \{0,1\}$). Additionally, we make the simplifying assumption that the reports correspond to local decisions made by the nodes about the system state. Specifically, we indicate by $r_{ij} \in \{0,1\}, i = 1 \dots n, j = 1 \dots m$ the report sent by node $i$ regarding the $j$-th component of $s^m$.

\begin{figure}[t!]
\centering
    \includegraphics[width=0.47\textwidth]{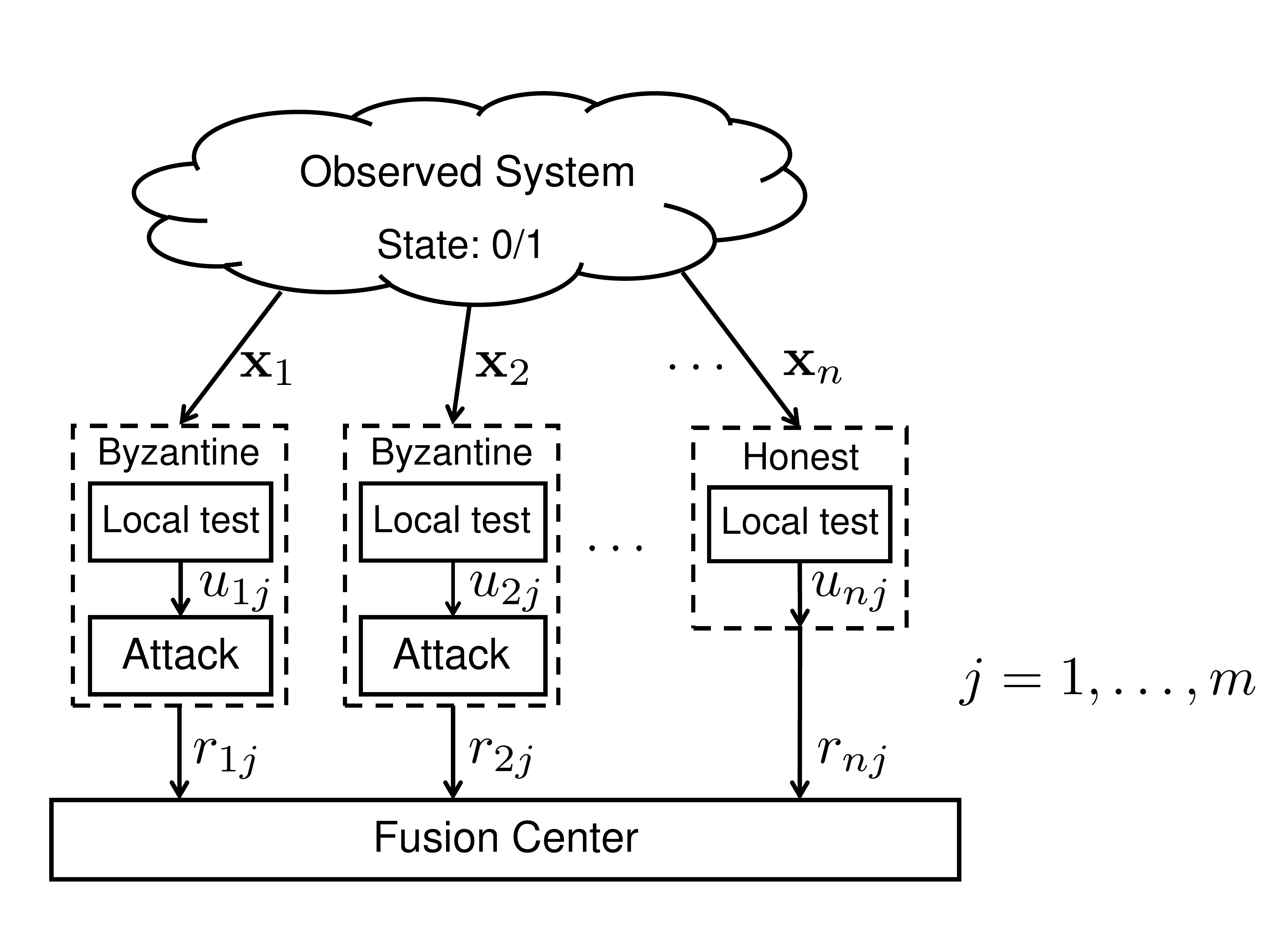}
    \caption{Sketch of the adversarial decision fusion scheme.}
    \label{fig.setup}
\end{figure}

Decision fusion must be carried out in an adversarial setting, that is by taking into account the possibility that some of the nodes malevolently alter their reports to induce a decision error. This is a recurrent situation in many scenarios where a decision error results in a profit for the nodes (see \cite{Vemp13} for a general introduction to this topic). To be specific, we assume that the nodes do not know the exact state of the system, so they estimate it based on the observation vectors ${\bf x}_i$'s. Let us denote with $u_{ij}$ the binary decision made by node $i$ regarding the $j$-th component of $s^m$. While for honest nodes $r_{ij} = u_{ij}$, malicious nodes flip $u_{ij}$ with a certain probability $P_{mal}$, so that $u_{ij} \ne r_{ij}$ with probability $P_{mal}$. Hereafter, we assume that the same probability $P_{mal}$ is used by all malicious nodes and for all the components of $s^m$, that is $P_{mal}$ does not depend either on $i$\footnote{Of course, assuming that the $i$-th node is a Byzantine, otherwise we obviously have $P_{mal} = 0$.} or $j$.
A pictorial representation of the setup analyzed in this paper is given in Figure \ref{fig.setup}.

\subsection{Prior work}

In a simplified and well studied version of the problem, the FC makes its decision on $s_j$ by looking only at the corresponding reports, i.e. $(r_{1,j}, r_{2,j} \dots r_{n,j})$. This is a reasonable assumption in some applications, e.g., when the components of $s^m$ correspond to the state of the observed system at different time instants, and a decision must be made as soon as the reports regarding the state at time $j$ are available. In the absence of Byzantines, the optimum way to combine the local decisions according to the Bayesian approach has been determined in \cite{OptFusion,Var97}, and is known as Chair-Varshney rule. If local error probabilities are symmetric and equal across the nodes, Chair-Varshney rule corresponds to simple majority-bases decision.

When Byzantines are present, the application of Chair-Varshney rule requires that the position of the byzantine nodes is known, along with the flipping probability $P_{mal}$, an information that is rarely available, thus forcing the FC to adopt suboptimal fusion strategies. In \cite{Mar09}, decision fusion is framed into a Neyman-Pearson setup and the asymptotic performance of the system when the number of nodes tends to infinity are analyzed as a function of the percentage of corrupted reports. As a result, the fraction of Byzantines impeding any correct decision is determined. Another noticeable aspect of \cite{Mar09} is that the Byzantines are assumed to cooperate among them to infer the exact status of the system and corrupt their reports accordingly. The analysis carried out in \cite{Mar09} is extended in \cite{Raw11}, where the interplay between the strategy adopted by the Byzantines to attack the system and the fusion rule adopted by FC together with the local decision strategy used by honest nodes is modeled as a zero sum game, whose payoff is either the overall error probability or the divergence between the probability mass function (pmf) of the observed reports under the hypothesis that $s = 0 $ and $s = 1$. Even in \cite{Raw11}, the authors determine the minimum fraction of Byzantines impeding any correct decision with both cooperative and noncooperative Byzantines.

Better results can be obtained if the FC collects all the reports and estimate the state vector as a whole. In a cognitive radio scenario, for instance, this corresponds to decide about spectrum occupancy over an entire time window, or, more realistically, to jointly decide about the state of the spectrum at different frequency slots. As an example, the FC may try to identify the malicious nodes by measuring the similarity (or dissimilarity) between the submitted reports and use such an estimate to ignore the reports coming from suspect nodes in the decision fusion process. Such an approach, which is usually referred to as Byzantine {\em isolation} \cite{Vemp13}, is adopted in \cite{Raw11}. According to such a work, all the components of the state vector are analyzed to assign to each node a reputation measure which is eventually used to isolate the byzantine nodes, which are identified as the nodes with a low reputation. A different isolation scheme based on adaptive learning is described in \cite{LearnByzantines}, where the observed behavior of the nodes is compared with the expected behavior of honest nodes. A peculiarity of this scheme is that it works even when the majority of the nodes are byzantine, but it requires very long state vectors to achieve good performances. Better performance can be  obtained if some additional knowledge about the Byzantine behavior is available, as in \cite{CDC}, where the knowledge about $P_{mal}$ and the number of Byzantines in the networks is exploited to develop a soft isolation scheme. As in \cite{KBKV13}, a game theoretic approach is used to determine the optimum strategies for the Byzantines and the FC. This corresponds to determining the optimum value of $P_{mal}$ for the Byzantines and the value of some internal parameters of the isolation scheme for the FC. As in that work,
it turns out that setting $P_{mal} = 1$ is a dominant strategy for the Byzantines.
Tolerant schemes which mitigate the impact of Byzantines in the decision, instead of removing them, have also been proposed, like in \cite{tolerant_scheme}, where the reports are weighted differently according to the reputation score of the nodes.

\subsection{Contribution}
\label{subsec.contrib}

Our research starts from the observation that the knowledge of $P_{mal}$ and the probability distribution of Byzantines across the network would allow the derivation of the optimum decision fusion rule, thus permitting to the FC to obtain the best achievable performance. We also argue that in the presence of such an information discarding the reports received from suspect nodes is not necessarily the optimum strategy, since such reports may still convey some useful information about the status of the system. This is the case, for instance, when $P_{mal} = 1$. If the FC knows the identity of byzantine nodes, in fact, it only needs to flip the reports received from such nodes to cancel the Byzantines' attack. In this sense, the method proposed in \cite{CDC} is highly suboptimal, since it does not fully exploit the knowledge of Byzantine distribution and $P_{mal}$.

As a first contribution, we derive the optimum decision fusion rule when the FC knows both the probability distribution of Byzantines and $P_{mal}$. Our analysis goes along a line which is similar to that used in \cite{OptFusion} to derive the Chair-Varshney optimal fusion rule. As a matter of fact, by knowing $P_{mal}$ and assuming that the probability that a node is Byzantine is fixed and independent on the other nodes, the Chair-Varshney rule can be easily extended to take into account the presence of Byzantines. In contrast to \cite{OptFusion}, however, the optimal fusion rule we derive in this paper, makes a joint decision on the whole sequence of states hence permitting to improve the decision accuracy. In addition, the analysis is not limited to the case of independently distributed
Byzantines. As an additional contribution, we describe an efficient implementation of the optimum fusion strategy based on dynamic programming.

Starting from the optimum decision fusion rule, the second contribution of this work focuses on the a-priori knowledge required to implement the optimum rule, namely $P_{mal}$ and the distribution of Byzantines across the network. In order to cope with the lack of knowledge regarding $P_{mal}$, we introduce a game-theoretic approach according to which the FC arbitrarily sets the value of $P_{mal}$ to a guessed value $P_{mal}^{FC}$ and uses such a value within the optimum fusion rule. At the same time, the Byzantines choose the value of $P_{mal}$ so to maximize the error probability, without knowing the value of $P_{mal}^{FC}$ used by the fusion center. The payoff is defined as the overall error probability, with the FC aiming at minimizing it, while the goal of the Byzantines is to maximize it. Having defined the game, we use numerical simulations to derive the existence of equilibrium points, which then identify the optimum behavior for both the FC and the Byzantines in a game-theoretic sense. While the adoption of a game-theoretic framework to model decision fusion in the presence of Byzantines has been used before, its adoption as an alternative to optimum decision fusion in the absence of precise information about Byzantines’ behavior is a novel contribution of this work. With regard to the knowledge that the FC has about the distribution of Byzantines, we consider several cases, ranging from a maximum entropy scenario in which the uncertainty about the distribution of Byzantines is maximum, through a more favorable situation in which the FC knows the exact number of Byzantines present in the network.

As a last contribution, we use numerical simulations to get more insights into the optimum strategies at the equilibrium and the achievable performance under various settings. The simulations show that in all the analyzed cases, the performance at the equilibrium outperform those obtained in previous works (specifically in \cite{Raw11, CDC}). Simulation results also confirm the intuition that, in some instances, it is preferable for the Byzantines to minimize the mutual information between the status of the observed system and the reports submitted to the FC, rather than always flipping the decision made by the local nodes as it is often assumed in previous works. This is especially true when the length of the observed sequence and the available information about the Byzantine distribution allow a good identification of byzantine nodes.

The rest of this paper is organized as follows. In Section \ref{sec.OptFus}, we derive the optimum fusion rule under different assumptions on the distribution of Byzantines. In Section \ref{sec.DP}, we propose an efficient implementation of the optimum fusion rule based on dynamic programming. In Section \ref{sec.GT}, we introduce the game-theoretic framework modeling the interplay between the Byzantines and the FC. In Section \ref{sec.simul}, we present simulations results and discuss optimum attacking and fusion strategies in various settings. We conclude the paper in Section \ref{sec.conc} with some final remarks.

\section{Optimum fusion rule}
\label{sec.OptFus}

In the rest of the paper, we will use capital letters to denote random variables and lowercase letters for their instantiations. Given a random variable $X$, we indicate with $P_X(x)$ its probability mass function (pmf). Whenever the random variable the pmf refers to is clear from the context, we will use the notation $P(x)$ as a shorthand for $P_X(x)$.

With the above notation in mind, we let $S^m = (S_1, S_2 \dots S_m)$ indicate a sequence of independent and identically distributed (i.i.d.) random variables indicating the state of the system. We assume that all states are equiprobable, that is $P_{S_j}(0) = P_{S_j}(1) = 0.5$.
We denote by $U_{ij} \in \{0,1\}$ the local decision made by node $i$ about $S_j$. We exclude any interaction between the nodes and assume that $U_{ij}$'s are conditionally independent for a fixed status of the system. This is equivalent to assuming that the local decision errors are i.i.d.

With regard to the position of the Byzantines, let $A^n = (A_1 \dots A_n)$ be a binary random sequence in which $A_i = 0$ (res. $A_i = 1$) if node $i$ is honest (res. byzantine). The probability that the distribution of Byzantines across the nodes is $a^n$ is indicated by $P_{A^n}(a^n)$ or  simply $P(a^n)$.

Finally, we let ${\bf R} = \{R_{ij}\}, ~ i = 1 \dots n, j = 1 \dots m$ be a random matrix with all the reports received by the fusion center, accordingly, we denote by ${\bf r} = \{r_{ij}\}$ a specific instantiation of ${\bf R}$. As stated before, $R_{ij} = U_{ij}$ for honest nodes, while $P(R_{ij} \ne U_{ij}) = P_{mal}$ for byzantine nodes. Byzantine nodes flip the local decisions $U_{ij}$ independently of each other with equal probabilities, so that their action can be modeled as a number of independent binary symmetric channels with crossover probability $P_{mal}$.

We are now ready to derive the optimum decision rule at the FC. Given the received reports ${\bf r}$ and by adopting a maximum a posteriori probability criterion, the optimum decision rule minimizing\ the error probability $P_e$ can be written as:

\begin{equation}
s^{m,*} = \arg\max_{s^m} P(s^m | {\bf r}).
\label{eq.map}
\end{equation}

By applying Bayes rule and exploiting the fact that all state sequences are equiprobable we obtain:
%
\begin{equation}
s^{m,*} =  \arg\max_{s^m} P({\bf r} | s^m ).
\label{eq.ML}
\end{equation}
In order to go on, we condition $P({\bf r} | s^m)$ to the knowledge of $a^n$ and then average over all possible $a^n$:

\begin{align}
s^{m,*}  
= & \arg\max_{s^m} \sum_{a^n} P({\bf r} | a^n, s^m) P(a^n) \label{eq.pseudoML}\\
= & \arg\max_{s^m} \sum_{a^n} \bigg(\prod_{i=1}^n P({\bf r}_i | a_i, s^m )\bigg) P(a^n)\label{eq.pseudoML_2}\\
= & \arg\max_{s^m} \sum_{a^n} \bigg(\prod_{i=1}^n \prod_{j=1}^m P(r_{ij} | a_i, s_j )\bigg) P(a^n),\label{eq.pseudoML_3}
\end{align}

where ${\bf r}_i$ indicates the $i$-th row of ${\bf r}$. In \eqref{eq.pseudoML_2} we exploited the fact that, given $a^n$ and $s^m$, the reports sent by the nodes are independent of each other, while \eqref{eq.pseudoML_3} derives from the observation that each report depends only on the corresponding element of the state sequence.


We now consider the case in which the probability of a local decision error, say $\varepsilon$, is the same regardless of the system status, that is $\varepsilon = Pr(U_{ij} \neq S_j|S_j = s_j)$, $s_j = 0,1$. For a honest node, such a probability corresponds to the probability that the report received by the FC does not correspond to the system status. This is not the case for byzantine nodes, for which the probability $\delta$ that the FC receives a wrong report is $\delta = \varepsilon (1 - P_{mal}) + (1 - \varepsilon)P_{mal}$.

According to the above setting, the nodes can be modeled as binary symmetric channels, whose input corresponds to the system status and for which the crossover probability is equal to $\varepsilon$ for the honest nodes and $\delta$ for the Byzantines. With regard to $\varepsilon$, it is reasonable to assume that such a value is known to the fusion center, since it depends on the characteristics of the channel through which the nodes observe the system and the local decision rule adopted by the nodes. The value of $\delta$ depends on the value of  $P_{mal}$ which is chosen by the Byzantines and then is not generally known to the FC. As we outlined in Section \ref{subsec.contrib}, we will first derive the optimum fusion rule assuming that $P_{mal}$ is known and then relax this assumption by modeling the problem in a game-theoretic framework (see Section \ref{sec.GT}).

From \eqref{eq.pseudoML_3}, the optimum decision rule can be written:
\begin{align}
\label{eq.pseudoML_symm}
s^{m,*} =  \arg\max_{s^m} & \sum_{a^n} \bigg(\prod_{i:a_i = 0}  (1-\varepsilon)^{m_{eq}(i)} \varepsilon^{m-m_{eq}(i)} \bigg. \\ & \hspace{0.5cm} \bigg. \prod_{i:a_i = 1} (1-\delta)^{m_{eq}(i)} \delta^{m-m_{eq}(i)} \bigg) P(a^n),\nonumber
\end{align}
where $m_{eq}(i)$ is the number of $j$'s for which $r_{ij} = s_j$. To go on we need to make some assumptions on the distribution of Byzantines $P(a^n)$.

\subsection{Unconstrained maximum entropy distribution}
\label{sec.MaxEnt}

As a worst case situation, we may consider that the FC has no a-priori information about the distribution of Byzantines. This corresponds to maximizing the entropy of $A^n$, i.e. to assuming that all sequences $a^n$ are equiprobable, $P(a^n) = 1/2^n$. In this case, the random variables $A_i$ are independent of each other and we have $P_{A_i}(0) = P_{A_i}(1) = 1/2$. It is easy to argue that in this case the Byzantines may impede any meaningful decision at the FC. To see why, let us assume that the Byzantines decide to use $P_{mal}=1$. With this choice, the mutual information between the vector state $S^m$ and ${\bf R}$ is zero and so any decision made by the FC center would be equivalent to {\em guessing} the state of the system by flipping a coin. The above observation is consistent with previous works in which it is usually assumed that the probability that a node is Byzantine or the overall fraction of Byzantines is lower than 0.5, since otherwise the Byzantines would always succeed to blind the FC \cite{Vemp13}.

\subsection{Constrained maximum entropy distributions}
\label{sec.ConstrMaxEnt}

A second possibility consists in maximizing the entropy of $A^n$ subject to a constraint which corresponds to the a-priori information available to the fusions center. We consider two cases. In the first one the FC knows the expected value of the number of Byzantines present in the network, in the second case, the FC knows only that the number of Byzantines is lower than $n/2$. In the following, we let $N_B$ indicate the number of Byzantines present in the network.

\subsubsection{Maximum entropy for a given $E[N_B]$}
\label{subsec.fixed_mean}

Let $\alpha = E[N_B]/n$ indicate the fraction of Byzantines nodes in the network. In order to determine the distribution $P{(a^n})$ which maximizes $H(A^n)$ subject to $\alpha$, we observe that $E[N_B] = E[\sum_i A_i] = \sum_i E[A_i] = \sum_i \mu_{A_i}$, where $\mu_{A_i}$ indicates the expected value of $A_i$. In order to determine the maximum entropy distribution constrained to $E[N_B] = \alpha n$, we need to solve the following problem:
\begin{equation}
\max_{P(a^n): \sum_i \mu_{A_i} = n\alpha} H(A^n).
\label{eq.MaxEntConstr}
\end{equation}
We now show that the solution to the above maximization problem is obtained by letting the $A_i$'s to be i.i.d. random variables with $\mu_{A_i} = \alpha$. We have:
\begin{equation}
    H(A^n) \le \sum_i H(A_i) = \sum_i h(\mu_{A_i}),
\label{eq.Crule}
\end{equation}
where $h(\mu_{A_i})$ denotes the binary entropy function\footnote{For any $p \le 1$ we have: $h(p) = p \log_2 p + (1-p) \log_2(1-p)$.} and where the last equality derives from the observation that for a binary random variable $A$, $\mu_A = P_A(1)$. We also observe that equality holds if and only if the random variables $A_i$'s are independent. To maximize the rightmost term in equation (\ref{eq.Crule}) subject to $\sum_i \mu_{A_i} = n \alpha$, we observe that the binary entropy is a concave function \cite{CandT}, and hence the maximum of the sum is obtained when all $\mu_{A_i}$'s are equal, that is when $\mu_{A_i} = \alpha$.

In summary, the maximum entropy case with known average number of Byzantines, corresponds to assuming i.i.d. node states for which the probability of being malicious is constant and known to the FC\footnote{Sometimes this scenario is referred to as Clairvoyant case \cite{Raw11}.}. We also observe that when $\alpha = 0.5$, we go back to the unconstrained maximum entropy case discussed in the previous section.

Let us assume, then, that $A_i$'s are Bernoulli random variables with parameter $\alpha$, i,.e., $P_{A_i}(1) = \alpha$, $\forall i$. In this way, the number of Byzantines in the network is a random variable with a binomial distribution. In particular, we have $P(a^n) = \prod_i P(a_i)$, and hence \eqref{eq.pseudoML_2} can be rewritten as:
\begin{equation}
s^{m,*} = \arg\max_{s^m} \sum_{a^n} \bigg(\prod_{i=1}^n P({\bf r}_i | a_i, s^m )P(a_i) \bigg).
\label{eq.factorization}
\end{equation}

The expression in round brackets corresponds to a factorization of $P({\bf r}, a^n | s^m )$.
If we look at that expression as a function of $a^n$, it is a product of marginal functions.
By exploiting the distributivity of the product with respect to the sum
we can rewrite \eqref{eq.factorization} as follows
\begin{equation}
s^{m,*} =  \arg\max_{s^m} \prod_{i=1}^{n} \bigg(\sum_{a_i \in \{0,1\}} P({\bf r}_i| a_i, s^m) P(a_i) \bigg),
\label{eq.pseudoML_Random}
\end{equation}
which can be computed more efficiently, especially for large $n$.
The expression in \eqref{eq.pseudoML_Random} can also be derived directly from \eqref{eq.ML} by exploiting first the independence of the reports and then applying the law of total probability.
By reasoning as we did to derive \eqref{eq.pseudoML_symm}, the to-be-maximized expression for the case of symmetric error probabilities at the nodes becomes
\begin{align}
s^{m,*} = & \arg\max_{s^m} \prod_{i=1}^{n} \left[(1-\alpha)(1-\varepsilon)^{m_{eq}(i)} \varepsilon^{m-m_{eq}(i)}\right.\nonumber\\
 & \hspace{3cm} \left.+ \alpha(1-\delta)^{m_{eq}(i)} \delta^{m-m_{eq}(i)}\right].
 \label{eq.pseudoML_Random_2}
\end{align}

Due to the independence of node states, the complexity of the above maximization problem grows only linearly with $n$, while it is exponential with respect to $m$, since it requires the evaluation of the to-be-minimized function for all possible sequence $s^m$. For this reason, the optimal fusion strategy can be adopted only when the length of the state sequence is limited.

\subsubsection{Maximum entropy with $N_B < n/2$}
\label{subsec.less_half}

As a second possibility, we assume that the FC knows only that the number of Byzantines is lower than $n/2$. As already observed in previous works \cite{Vemp13,WSNMatta,Raw11}, and as it is easily arguable, when the number of Byzantines exceeds the number of honest nodes no meaningful decision can be made. Then it makes sense for the FC to assume that $N_B < n/2$, since if this is not the case, no correct decision can be made anyhow. Under this assumption, the maximum entropy distribution is the one which assigns exactly the same probability to all the sequences $a^n$ for which $\sum_i a_i < n/2$. To derive the optimum decision fusion strategy in this setting, let $\mathcal{I}$ be the indexing set $\{1,2,...,n\}$.  We denote with $\mathcal{I}_k$ the set of all the possible $k$-subsets of $\mathcal{I}$. Let $I \in \mathcal{I}_k$ be a random variable with the indexes of the byzantine nodes, a node $i$ being byzantine if $i \in I$, honest otherwise. We this notation, we can rewrite \eqref{eq.pseudoML} as
\begin{align}
s^{m,*} = \arg\max_{s^m}  \sum_{k=0}^{\lfloor n/2 \rfloor }\sum_{I \in \mathcal{I}_{k}} P({\bf r} | I, s^m ) p(s^m),\label{eq.ML_less_thann2_general}
\end{align}
where we have omitted the term $P(I)$ (or equivalently $P(a^m)$) since all the sequences for which $N_B < n/2$ have the same probability. In the case of symmetric local error probabilities, \eqref{eq.ML_less_thann2_general}
takes the following form:
\begin{align}
s^{m,*} =  \arg\max_{s^m} &  \sum_{k=0}^{\lfloor n/2 \rfloor} \sum_{I \in \mathcal{I}_{k}} \bigg(\prod_{i \in I} (1-\delta)^{m_{eq}(i)} \delta^{m-m_{eq}(i)}\bigg.\nonumber\\
  & \hspace{1cm} \bigg.\prod_{i \in \mathcal{I} \setminus I} (1-\varepsilon)^{m_{eq}(i)} \varepsilon^{m-m_{eq}(i)}\bigg).
  \label{eq.ML_less_thann2_2}
\end{align}
A problem with \eqref{eq.ML_less_thann2_2} is the complexity of the inner summation, which grows exponentially with $n$ (especially for values of $k$ close to $n/2$). Together with the maximization over all possible $s^m$, this results in a doubly exponential complexity, making the direct implementation of \eqref{eq.ML_less_thann2_2} problematic. In Section \ref{sec.DP}, we introduce an efficient algorithm based on dynamic programming which reduces the computational complexity of the maximization in \eqref{eq.ML_less_thann2_2}.

We conclude by stressing an important difference between the case considered in this subsection and the maximum entropy case with fixed $E[N_B]$. On one hand, the average number of Byzantines in the fixed average case may be lower than in the present case. On the other hand, such a setting, does not guarantee that the number of Byzantines is always lower than the number of honest nodes, as in the case analyzed in this subsection. This observation will be crucial to explain some of the results that we will present later on in the paper.

\subsection{Fixed number of Byzantines}
\label{sec.OF_DETstates}

The final setting we are going to analyze assumes that the fusion center knows the exact number of Byzantines, say $n_B$. This is a more favorable situation with respect to those addressed so far. The derivation of the optimum decision fusion rule stems from the observation that, in this case, $P(a^n) \ne 0$ only for the sequence for which $\sum_i a_i = n_B$. For such sequences, $P(a^n)$ is constant and equal to $\binom{n}{n_B}^{-1}$. By using the same notation used in the previous section, the optimum fusion rules, then, is:
\begin{align}
s^{m,*} = \arg\max_{s^m}  \sum_{I \in \mathcal{I}_{n_B}} P({\bf r} | I, s^m ) p(s^m),\label{eq.ML_Determ}
\end{align}
which reduces to
\begin{align}
s^{m,*} =  \arg\max_{s^m} &  \sum_{I \in \mathcal{I}_{n_B}} \bigg(\prod_{i \in I} (1-\delta)^{m_{eq}(i)} \delta^{m-m_{eq}(i)}\bigg.\nonumber\\
  & \hspace{1cm} \bigg.\prod_{i \in \mathcal{I} \setminus I} (1-\varepsilon)^{m_{eq}(i)} \varepsilon^{m-m_{eq}(i)}\bigg),
  \label{eq.ML_Determ_2}
\end{align}
in the case of equal local error probabilities. With regard to computational complexity, even if the summation over all possible number of Byzantines is no more present, the direct implementation of \eqref{eq.ML_Determ_2} is still very complex due to the exponential dependence of the cardinality of $\mathcal{I}_{n_B}$ with respect to $n$.

\section{An efficient implementation based on dynamic programming}
\label{sec.DP}

The computational complexity of a direct implementation of \eqref{eq.ML_less_thann2_2} and \eqref{eq.ML_Determ_2} hinders the derivation of the optimum decision fusion rule for large size networks. Specifically, the problem with \eqref{eq.ML_less_thann2_2} and \eqref{eq.ML_Determ_2} is the exponential number of terms of the summation over $\mathcal{I}_k$ ($\mathcal{I}_{n_B}$ in \eqref{eq.ML_Determ_2}). In this section, we show that an efficient implementation of such summations is possible based on Dynamic Programming (DP) \cite{DynamicProgramming}.

Dynamic programming is an optimization strategy which allows to solve complex problems by transforming them into subproblems and by taking advantage of the subproblems overlap in order to reduce the number of operations . When facing with complex recursive problems, by using dynamic programming we solve each different subproblem only once by storing the solution for subsequent use. If during the recursion the same subproblem is encountered again, the problem is not solved twice since its solution is already available. Such a re-use of previously solved subproblems is often referred in literature as memorization algorithm \cite{DynamicProgramming}. Intuitively, DP allows to reduce the complexity of problems with a structure, such that the solutions of the same subproblems can be reused many times.

\begin{figure}[h!]
\centering
    \includegraphics[width=0.8\columnwidth]{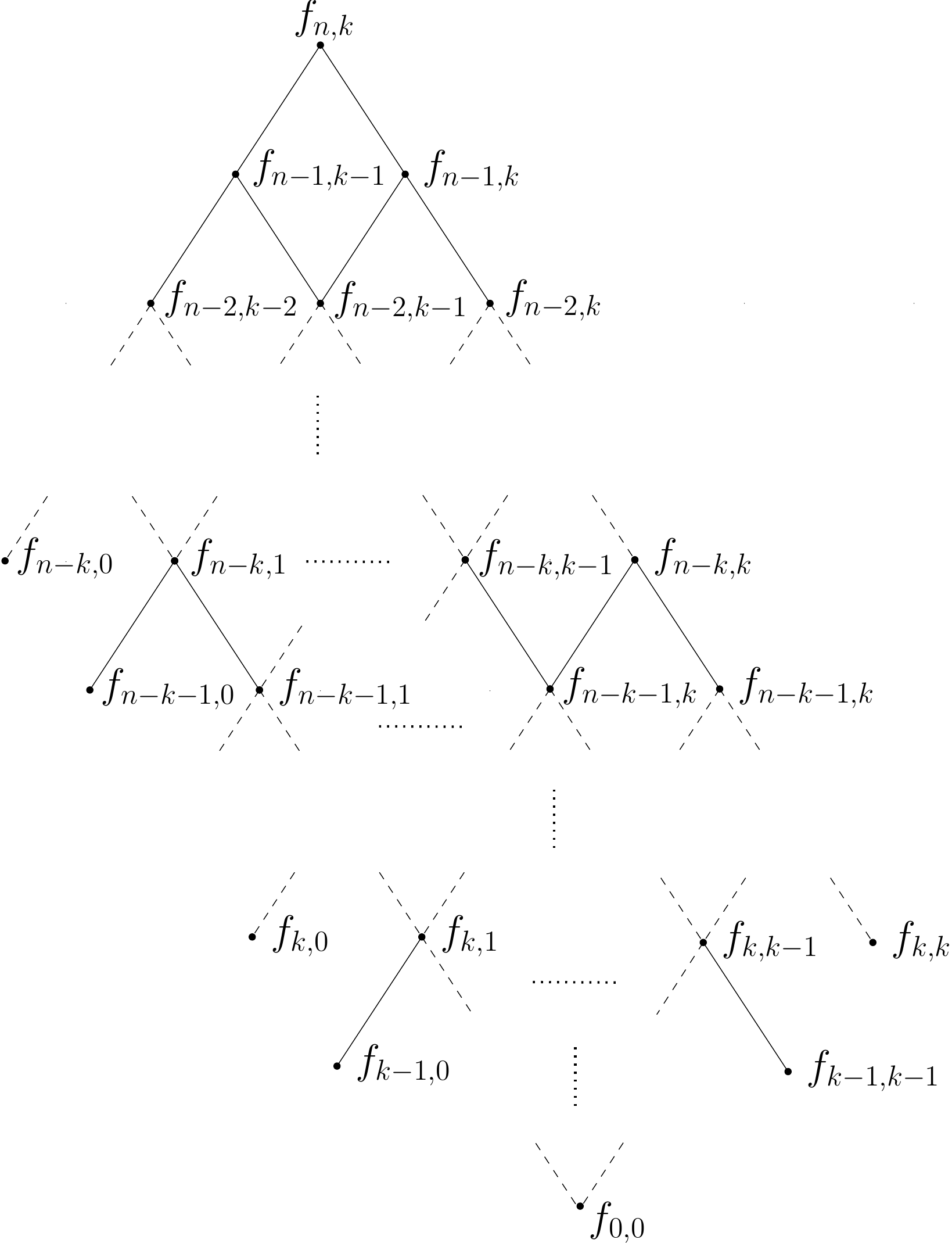}\vspace{0.4cm}
    \caption{Efficient implementation of the function in \eqref{function_f_n_k} based on dynamic programming. The figure depicts the tree with the iterations for the case $k < n-k$.}
    \label{fig.OptRule_DPcomplexity}
\end{figure}

We now apply dynamic programming to reduce the complexity of our problem.
Let us focus on a fixed $k$ (and $n$) and let us define the function $f_{n,k}$ as follows:
\begin{align}
\label{function_f_n_k}
f_{n,k} =  & \sum_{I \in \mathcal{I}_{k}} \bigg(\prod_{i \in I} (1-\delta)^{m_{eq}(i)} \delta^{m-m_{eq}(i)}\bigg.\nonumber\\
  & \hspace{1cm} \bigg.\prod_{i \in \mathcal{I} \setminus I} (1-\varepsilon)^{m_{eq}(i)} \varepsilon^{m-m_{eq}(i)}\bigg).
\end{align}
%
%
By focusing on node $i$, there are some configurations $I \in \mathcal{I}_{k}$ for which such a node belongs to $I$, while for others the node belongs to the complementary set $\mathcal{I} \setminus I$. Let us define $b(i) = (1-\delta)^{m_{eq}(i)} \delta^{m-m_{eq}(i)}$ and $h(i) = (1-\varepsilon)^{m_{eq}(i)} \varepsilon^{m-m_{eq}(i)}$, which are the two contributions that node $i$ can provide to each term of the sum, depending on whether it belongs to $\mathcal{I}$ or $\mathcal{I} \setminus I$. Let us focus on the first indexed node. By exploiting the distributivity of the product with respect to the sum, expression \eqref{function_f_n_k} can be rewritten in a recursive manner as:
\begin{align}
f_{n,k} =  b(1) f_{n-1,k-1} + h(1) f_{n-1,k}.
\end{align}
By focusing on the second node, we can iterate on $f(n-1,k-1)$  and  $f(n-1,k)$, getting:
\begin{align}
f_{n-1,k-1} =  b(2) f_{n-2,k-2} +  h(2) f_{n-2,k-1},
\label{eq.sub1}
\end{align}
and
\begin{align}
\label{eq.sub2}
f_{n-1,k}= b(2) f_{n-2,k-1} + h(2) f_{n-2,k}.
\end{align}
We notice that the subfunction $f_{n-2,k-1}$ appears in both \eqref{eq.sub1} and \eqref{eq.sub2} and then it can be computed only once. The procedure can be iterated for each subfunction until we reach a subfunction whose value can be computed in closed form, that is:  $f_{r,r} = \prod_{i =n-r + 1}^n b(i)$ and $f_{r,0} = \prod_{i =n-r + 1}^n h(i)$, for some $r \le k$.
By applying the memorization strategy typical of dynamic programming, the number of required computations is given by the number of nodes in the tree depicted in Figure \ref{fig.OptRule_DPcomplexity}, where the leaves correspond to the terms computable in closed form\footnote{The figure refers to the case $k < n-k$, which is always the case in our setup since $k < \lfloor n/2 \rfloor$.}. By observing that the number of the nodes of the tree is $k(k + 1)/2 + k(n - k - k) + k(k+1)/2 = k(n - k +1)$, we conclude that the number of operations is reduced from $\binom{n}{k}$ to $k(n - k +1)$, which corresponds to a quadratic complexity instead of an exponential one.
%
%

\section{Decision fusion with Byzantines game}
\label{sec.GT}

The optimum decision fusion rules derived in Section \ref{sec.OptFus} assume that the FC knows the attacking strategy adopted by the Byzantines, which in the simplified case studied in this paper corresponds to knowing $P_{mal}$. By knowing $P_{mal}$, in fact, the FC can calculate the value of $\delta$ used in equations (\ref{eq.pseudoML_symm}), (\ref{eq.pseudoML_Random_2}), (\ref{eq.ML_less_thann2_2}) and (\ref{eq.ML_Determ_2}), and hence implement the optimum fusion rule. In previous works, as in \cite{Rawatconf, Raw11}, it is often conjectured that $P_{mal} = 1$. In some particular settings, as the ones addressed in \cite{KBKV13, CDC}, it has been shown that this choice permits to the Byzantines to maximize the error probability at the fusion center. Such an argument, however, does not necessarily hold when the fusion center can localize the byzantine nodes with good accuracy and when it knows that the byzantine nodes always flip the local decision. In such a case, in fact, the FC can revert the action of the Byzantines by simply inverting the reports received from such nodes, as it is implicitly done by the optimal fusion rules derived in the previous section. In such a situation, it is easy to argue that it is better for the Byzantines to let $P_{mal} = 0.5$ since in this way the mutual information between the system status and the reports received from the byzantine nodes is equal to zero. In general, the byzantine nodes must face the following dilemma: is it better to try to force the FC to make a wrong decision by letting $P_{mal} = 1$ and run the risk that if their location in the network is detected the FC receives some useful information from the corrupted reports, or erase the information that the FC receives from the attacked nodes by reducing to zero the mutual information between the corrupted reports and $S^m$ ?

Given the above discussion, it is clear that the FC can not assume that the Byzantines use $P_{mal} = 1$, hence making the actual implementation of the optimum decision fusion rule impossible.

In order to exit this apparent deadlock, we propose to model the struggle between the Byzantines and the FC as a two-player, zero-sum, strategic game, whose equilibrium defines the optimum choices for the FC and the Byzantines.

\subsection{Game theory in a nutshell}

A 2-player game is defined as a 4-uple $G(\SS_1,\SS_2,v_1, v_2)$, where $\SS_1 = \{z_{1,1} \dots z_{1,n_1}\}$ and $\SS_2 = \{z_{2,1} \dots z_{2,n_2}\}$ are the set of actions (usually called strategies) the first and the second player can choose from, and $v_l(z_{1,i}, z_{2,j}), l= 1,2$, is the payoff of the game for player $l$, when the first player chooses the strategy $z_{1,i}$ and the second chooses $z_{2,j}$. A pair of strategies $(z_{1,i}, z_{2,j})$ is called a profile. When $v_1(z_{1,i}, z_{2,j}) + v_2(z_{1,i}, z_{2,j}) = 0$, the game is said to be a zero-sum game. In the set-up adopted in this paper, $\SS_1$, $\SS_2$ and the payoff functions are assumed to be known to the two players. In addition, we assume that the players choose their strategies before starting the game without knowing the  strategy chosen by the other player (strategic game).

A common goal in game theory is to determine the existence of equilibrium points, i.e. profiles that in {\em some way} represent a {\em satisfactory} choice for both players \cite{Osb94}. The most famous equilibrium notion is due to Nash. Intuitively, a profile is a Nash equilibrium if each player does not have any interest in changing his choice assuming the other does not change his strategy. For the particular case of a 2-player game, a profile $(z_{1,i^*}, z_{2,j^*})$ is a Nash equilibrium if:
\begin{equation}
\begin{array}{ll}
    v_1((z_{1,i^*}, z_{2,j^*})) \ge v_1((z_{1,i}, z_{2,j^*})) & \forall z_{1,i} \in \SS_1\\
    v_2((z_{1,i^*}, z_{2,j^*})) \ge v_2((z_{1,i^*}, z_{2,j})) & \forall z_{2,j} \in \SS_2,
\end{array}
\label{eq.Nash}
\end{equation}
where for a zero-sum game $v_2 = -v_1$.

A stronger equilibrium notion is that of {\em dominant equilibrium}. A strategy is said to be strictly dominant for one player if it is the best strategy for the player, regardless of the strategy chosen by the other player. In many cases dominant strategies do not exist, however when one such strategy exists for one of the players, he will surely adopt it (at least under the assumption of  rational behavior). The other players, in turn, will choose their strategies anticipating that the first player will play the dominant strategy. As a consequence, in a two-player game, if a dominant strategy exists the players have only one rational choice called the only rationalizable equilibrium of the game \cite{ChenGames}. Games with the above property are called {\em dominance solvable} games.

The above definition assumes that the players deterministically choose one of the strategies in $\SS_i$ (pure strategy). A more flexible approach consists in letting each player choose a strategy with a certain probability. In this way, we introduce a new game in which the strategies available to the players are probability distributions over $\SS_i$'s. The payoff is redefined as the average payoff under the probability distributions chosen by the players. A probability distribution over $\SS_i$'s is said a mixed strategy for player $i$.  A central result of game theory \cite{Nash50} states that if we allow mixed strategies, then every game with a finite number of players and with a finite number of pure strategies for each player has at least one Nash equilibrium.

As anticipated, we model the interplay between the value of $P_{mal}$ adopted by the Byzantines and the value used by the FC in its attempt to implement the optimum fusion rule as game. For sake of clarity, in the following we indicate the flipping probability adopted by the Byzantines as $P_{mal}^B$, while we use the symbol $P_{mal}^{FC}$ to indicate the value adopted by the FC in its implementation of the optimum fusion rule. With the above ideas in mind, we introduce the Decision Fusion Game.

\begin{definition}
The $DF_{Byz}(\SS_{B}, \SS_{FC}, v)$ game is a two player, zero-sum, strategic, game played by the FC and the Byzantines (collectively acting as a single player), defined by the following strategies and payoff.
\begin{itemize}
\item{The sets of strategies the Byzantines and the FC can choose from are, respectively, the set of possible values of $P_{mal}^B$ and $P_{mal}^{FC}$:
\begin{eqnarray}
\SS_B = \{ P_{mal}^B \in [0,1]\};\nonumber \\
\SS_{FC} = \{ P_{mal}^{FC} \in [0,1]\}.
\label{eq.DFgameS}
\end{eqnarray}
}
\item{The payoff function is defined as the error probability at the FC, hereafter indicated as $P_e$. Of course the Byzantines aim at maximizing $P_e$, while the FC aims at minimizing it.}
\end{itemize}
\label{def.DFgame}
\end{definition}

Note that according to the definition of $DF_{Byz}$, the sets of strategies available to the FC and the Byzantines are continuous sets. In practice, however, continuous values can be replaced by a properly quantized version of $P_{mal}^B$ and $P_{mal}^{FC}$.

In the next section, we use numerical simulations to derive the equilibrium point of various versions of the game obtained by varying the probability distribution of byzantine nodes as detailed in Section \ref{sec.OptFus}. As we will see, while some versions of the game has a unique Nash (or even dominant) equilibrium point in pure strategies, in other cases, a Nash equilibrium exists only in mixed strategies.

\section{Simulation results and discussion}
\label{sec.simul}

In order to investigate the behavior of the $DF_{Byz}$ game for different setups and analyze the achievable performance when the FC adopts the optimum decision strategy with parameters tuned following a game-theoretic approach, we run extensive numerical simulations. The first goal of the simulations was to study the existence of an equilibrium point in pure or mixed strategies, and analyze the expected behavior of the FC and the Byzantines at the equilibrium. The second goal was to evaluate the payoff at the equilibrium as a measure of the best achievable performance of Decision Fusion in the presence of Byzantines. We then used such a value to compare the performance of the game-theoretic approach proposed in this paper with respect to previous works.

\subsection{Analysis of the equilibrium point of the $DF_{Byz}$ game}

As we said, the first goal of the simulations was to determine the existence of an equilibrium point for the $DF_{Byz}$ game. To do so we quantized the set of available strategies considering the following set of values: $P_{mal}^B \in \{0.5, 0.6, 0.7, 0.8, 0.9, 1\}$ and $P_{mal}^{FC} \in \{0.5, 0.6, 0.7, 0.8, 0.9, 1\}$. We restricted our analysis to values larger than or equal to $0.5$ since it is easily arguable that such values always lead to better performance for the Byzantines\footnote{By using a game-theoretic terminology, this is equivalent to say that the strategies corresponding to $P_{mal}^B < 0.5$ are dominated strategies and hence can be eliminated.} (in fact, when $P_{mal}^B = 0.5$ the mutual information between the system state and the reports sent to the FC is equal to zero). As to the choice of the quantization step, we set it to 0.1 to ease the description of the results we have got and speed up the simulations. Some exploratory test made with a smaller step gave similar results. We found that among all the parameters of the game, the value of $m$ has a major impact on the equilibrium point. The value of $m$, in fact, determines the ease with which the FC can localize the byzantine nodes, and hence plays a major role in determining the optimum attacking strategy for the Byzantines. For this reason, we split our analysis in two parts: the former refers to small values of $m$, the latter to medium values of $m$. Unfortunately,  the exponential growth of the complexity of the optimum decision fusion rule as a function of $m$ prevented us from running simulations with large values of $m$.

We run simulations to evaluate the payoff matrix of the game between the Byzantines and
the FC. To do so, we run 50,000 trials to compute the $P_e$ at each row of the matrix. In particular, for each
$P_{mal}^B$, we used the same 50,000 states to compute the $P_e$ for all $P_{mal}^{FC}$ strategies. In all the simulations, we let $P_{S_j}(0) = P_{S_j}(1) = 0.5$, $n=20$, and $\varepsilon=0.1$.

\subsubsection{Small $m$}

For the first set of simulations, we used a rather low value of $m$, namely $m = 4$. The other parameters of the game we set as follows: $n = 20$, $\varepsilon = 0.1$. With regard to the number of byzantine nodes present in the network we used $\alpha = \{0.3, 0.4, 0.45\}$ for the case of independent node states studied in Section \ref{subsec.fixed_mean}, and $n_B = \{6, 8, 9\}$ for the case of known number of Byzantines (Section \ref{sec.OF_DETstates}). Such values were chosen so that in both cases we have the same average number of Byzantines, thus easing the comparing between the two settings.

Tables \ref{tab.indip03m4} through \ref{tab.indip045m4} report the payoff for all the profiles resulting from the quantized values of $P_{mal}^B$ and $P_{mal}^{FC}$, for the case of independent node states (constrained maximum entropy distribution). In all the cases $P_{mal}^B = 1$ is a dominant strategy for the Byzantines, and the profile $(1,1)$ is the unique rationalizable equilibrium of the game. As expected, the error probability increases with the number of Byzantines.

\begin{table}[h]
\centering
\renewcommand{\arraystretch}{1.1}
\begin{tabular}{c| c| c| c| c| c| c|}
                       \hline
\multicolumn{1}{|c|}{\tiny{$P_{mal}^{B}$/$P_{mal}^{FC}$}} & 0.5   & 0.6   & 0.7   & 0.8   &0.9   &1.0   \\ \hline
\multicolumn{1}{|c|}{0.5}  &0.845 & 0.965 & 1.1 & 1.3 & 1.6 & 2.1\\ \hline
\multicolumn{1}{|c|}{0.6} &1.2 & 1.1 & 1.2 & 1.5e-3 & 1.8 & 2.6\\ \hline
\multicolumn{1}{|c|}{0.7}  &2.2 & 2.0 & 1.8 & 1.8e-3 & 2.1 & 3.7\\ \hline
\multicolumn{1}{|c|}{0.8} &5.4 & 5.1 & 5.0 & 5.0e-3 & 5.1 & 7.7\\ \hline
\multicolumn{1}{|c|}{0.9}  &16.2 & 16.1 & 16.5 & 16.4 & 16.0 & 19.1\\ \hline
\multicolumn{1}{|c|}{1.0} &43 & 43.1 & 46.9 & 46.8 & 41.6 & {\bf 34.9}\\ \hline
\end{tabular}

\caption{Payoff of the $DF_{Byz}$ game ($10^3 \times P_e$) with independent node states with $\alpha = 0.3$, $m= 4$, $n = 20$, $\varepsilon = 0.1$. The equilibrium point is highlighted in bold.}
\label{tab.indip03m4}
\end{table}

\begin{table}[h]
\centering
\renewcommand{\arraystretch}{1.1}
\begin{tabular}{c| c| c| c| c| c| c|}
                       \hline
\multicolumn{1}{|c|}{\tiny{$P_{mal}^{B}$/$P_{mal}^{FC}$}} & 0.5   & 0.6   & 0.7   & 0.8   &0.9   &1.0   \\ \hline
\multicolumn{1}{|c|}{0.5}  &0.33 &0.37  &0.44  &0.58 &0.73  &0.85 \\ \hline
\multicolumn{1}{|c|}{0.6}  &0.60 &0.54  &0.59  &0.70 &0.80  &1.14 \\ \hline
\multicolumn{1}{|c|}{0.7}  &1.38 &1.20  &1.19  &1.24 &1.29  &2.40 \\ \hline
\multicolumn{1}{|c|}{0.8}  &3.88 &3.56  &3.36  &3.31 &3.35  &6.03 \\ \hline
\multicolumn{1}{|c|}{0.9}  &9.93 &9.61  &9.57  &9.55 &9.54  &11.96 \\ \hline
\multicolumn{1}{|c|}{1.0}  &20.33 &20.98  &21.70  &21.90 &21.84  & {\bf 19.19} \\ \hline
\end{tabular}

\caption{Payoff of the $DF_{Byz}$ game ($10^2 \times P_e$) with independent node states with $\alpha = 0.4$, $m= 4$, $n = 20$, $\varepsilon = 0.1$. The equilibrium point is highlighted in bold.}
\label{tab.indip04m4}
\end{table}

\begin{table}[h]
\centering
\renewcommand{\arraystretch}{1.1}
\begin{tabular}{c| c| c| c| c| c| c|}
                       \hline
\multicolumn{1}{|c|}{\tiny{$P_{mal}^{B}$/$P_{mal}^{FC}$}} & 0.5   & 0.6   & 0.7   & 0.8   &0.9   &1.0   \\ \hline
\multicolumn{1}{|c|}{0.5}  &0.62 &0.69  &0.86  &1.34 &1.70  &1.57 \\ \hline
\multicolumn{1}{|c|}{0.6}  &1.23 &1.15  &1.26  &1.84 &2.18  &2.38 \\ \hline
\multicolumn{1}{|c|}{0.7}  &2.94 &2.64  &2.57  &3.00 &3.14  &5.33 \\ \hline
\multicolumn{1}{|c|}{0.8}  &7.89 &7.39  &7.03  &6.74 &6.81  &12.73 \\ \hline
\multicolumn{1}{|c|}{0.9}  &18.45 &17.94  &17.63  &17.08 &17.07  &22.78 \\ \hline
\multicolumn{1}{|c|}{1.0}  &34.39 &34.62  &34.84  &36.66 &36.61  & {\bf 33.14} \\ \hline
\end{tabular}

\caption{Payoff of the $DF_{Byz}$ game ($10^2 \times P_e$) with independent node states with $\alpha = 0.45$, $m= 4$, $n = 20$, $\varepsilon = 0.1$. The equilibrium point is highlighted in bold.}
\label{tab.indip045m4}
\end{table}

Tables \ref{tab.fixed6m4} through \ref{tab.fixed9m4} report the payoffs for the case of fixed number of Byzantines, respectively equal to 6, 8 and 9.

\begin{table}[h!]
\centering
\renewcommand{\arraystretch}{1.1}
\begin{tabular}{c| c| c| c| c| c| c|}
                       \hline
\multicolumn{1}{|c|}{\tiny{$P_{mal}^{B}$/$P_{mal}^{FC}$}} & 0.5   & 0.6   & 0.7   & 0.8   &0.9   &1.0   \\ \hline
\multicolumn{1}{|c|}{0.5}  & {\bf 3.80} & {\bf 3.80}  &4.60  &7.60 &12.0  &29.0 \\ \hline
\multicolumn{1}{|c|}{0.6}  &3.60 &3.45  &3.90  &5.20 &8.0  &17.0 \\ \hline
\multicolumn{1}{|c|}{0.7}  &3.45 &2.80  &2.80  &3.10 &4.40  &8.75 \\ \hline
\multicolumn{1}{|c|}{0.8}  &4.10 &2.85  &2.15  &2.05 &2.25  &3.25 \\ \hline
\multicolumn{1}{|c|}{0.9}  &3.55 &2.05  &1.40  &0.95 &0.70  &0.75 \\ \hline
\multicolumn{1}{|c|}{1.0}  &2.05 &0.90  &0.35  &0.15 &0.05  &0.05  \\ \hline
\end{tabular}

\caption{Payoff of the $DF_{Byz}$ game ($10^4 \times P_e$) with $n_B = 6$, $m= 4$, $n = 20$, $\varepsilon = 0.1$. The equilibrium point is highlighted in bold.}
\label{tab.fixed6m4}
\end{table}

\begin{table}[h!]
\centering
\renewcommand{\arraystretch}{1.1}
\begin{tabular}{c| c| c| c| c| c| c|}
                       \hline
\multicolumn{1}{|c|}{\tiny{$P_{mal}^{B}$/$P_{mal}^{FC}$}} & 0.5   & 0.6   & 0.7   & 0.8   &0.9   &1.0   \\ \hline
\multicolumn{1}{|c|}{0.5}  &1.2 &1.4  &1.9  &3.1 &6.3  &18.9 \\ \hline
\multicolumn{1}{|c|}{0.6}  &1.5 &1.4  &1.4  &2.0 &3.7  &10.0 \\ \hline
\multicolumn{1}{|c|}{0.7}  &1.4 &1.1  &0.945  &1.1 &1.7  &4.0 \\ \hline
\multicolumn{1}{|c|}{0.8}  &1.4 &0.95  &0.715  &0.58 &0.675  &1.2 \\ \hline
\multicolumn{1}{|c|}{0.9}  &2.1 &1.4  &0.995  &0.745 &0.71  &0.78 \\ \hline
\multicolumn{1}{|c|}{1.0}  &7.3 &5.7  &5.3  &3.7 &3.0  &2.9  \\ \hline
\end{tabular}

\caption{Payoff of the $DF_{Byz}$ game ($10^3 \times P_e$) with $n_B = 8$, $m= 4$, $n = 20$, $\varepsilon = 0.1$. No pure strategy equilibrium exists.}
\label{tab.fixed8m4}
\end{table}

\begin{table}[h!]
\centering
\renewcommand{\arraystretch}{1.1}
\begin{tabular}{c| c| c| c| c| c| c|}
                       \hline
\multicolumn{1}{|c|}{\tiny{$P_{mal}^{B}$/$P_{mal}^{FC}$}} & 0.5   & 0.6   & 0.7   & 0.8   &0.9   &1.0   \\ \hline
\multicolumn{1}{|c|}{0.5}  &0.22 &0.24  &0.33  &0.63 &1.41  &4.13 \\ \hline
\multicolumn{1}{|c|}{0.6}  &0.27 &0.24  &0.27  &0.41 &0.78  &2.03 \\ \hline
\multicolumn{1}{|c|}{0.7}  &0.32 &0.24  &0.23  &0.26 &0.37  &0.82 \\ \hline
\multicolumn{1}{|c|}{0.8}  &0.54 &0.45  &0.39  &0.36 &0.41  &0.59 \\ \hline
\multicolumn{1}{|c|}{0.9}  &2.04 &1.87  &1.76  &1.58 &1.56  &1.66 \\ \hline
\multicolumn{1}{|c|}{1.0}  &9.48 &8.76  &8.37  &6.72 &5.88  & {\bf 5.51} \\ \hline
\end{tabular}

\caption{Payoff of the $DF_{Byz}$ game ($10^2 \times P_e$) with $n_B = 9$, $m= 4$, $n = 20$, $\varepsilon = 0.1$. The equilibrium point is highlighted in bold.}
\label{tab.fixed9m4}
\end{table}

When $n_B = 6$, $P_{mal}^B = 0.5$ is a dominant strategy for the Byzantines, and the profile $(0.5, 0.5)$ is the unique rationalizable equilibrium of the game. This marks a significant difference with respect to the case of independent nodes, where the optimum strategy for the Byzantines was to let $P_{mal}^B = 1$. The reason behind the different behavior is that in the case of fixed number of nodes, the a-priori knowledge available at the FC is larger than in the case of independent nodes with the same average number of nodes. This additional information permits to the FC to localize the byzantine nodes, which now can not use $P_{mal}^B = 1$, since in this case they would still transmit some useful information to the FC. On the contrary, by letting $P_{mal}^B = 0.5$ the information received from the byzantine nodes is zero, hence making the task of the FC harder. When $n_B = 9$ (Table \ref{tab.fixed9m4}), the larger number of Byzantines makes the identification of malicious nodes more difficult and $P_{mal}^B = 1$ is again a dominant strategy, with the equilibrium of the game obtained at the profile (1,1). A somewhat intermediate situation is observed when $n_B = 8$ (Table \ref{tab.fixed8m4}). In this case, no equilibrium point exists (let alone a dominant strategy) if we consider pure strategies only. On the other hand, when mixed strategies are considered, the game has a unique Nash equilibrium for the strategies reported in Table \ref{tab.mixedNASHfixed8m4} (each row in the table gives the probability vector assigned to the quantized values of $P_{mal}$ by one of the players at the equilibrium). Interestingly the optimum strategy of the Byzantines corresponds to alternate playing $P_{mal}^{B} = 1 $ and $P_{mal}^{B} = 0.5 $, with intermediate probabilities. This confirms the necessity for the Byzantines to find a good trade-off between two alternative strategies: set to zero the information transmitted to the FC or try to push it towards a wrong decision. We also observe that the error probabilities at the equilibrium are always lower than those of the game with independent nodes. This is an expected result, since in the case of fixed nodes the FC has a better knowledge about the distribution of byzantine nodes.

\begin{table}[h!]
\centering
\renewcommand{\arraystretch}{1.4}
\begin{tabular}{c| c| c| c| c| c| c|}
                       \hline
\multicolumn{1}{|c|}{}& 0.5   & 0.6   & 0.7   & 0.8   &0.9   &1.0   \\ \hline
\multicolumn{1}{|c|}{$P(P_{mal}^{B})$ }  & 0.179 & 0 & 0 & 0 & 0 & 0.821 \\ \hline
\multicolumn{1}{|c|}{$P(P_{mal}^{FC})$ }  & 0 & 0  & 0  &0.844 & 0.156 &0 \\ \hline
\multicolumn{7}{|c|}{$P_e^* = 3.8e-4$} \\ \hline
\end{tabular}

\caption{Mixed equilibrium point for the $DF_{Byz}$ game with $n_B = 8$, $m= 4$, $n = 20$,  $\varepsilon = 0.1$. $P_e^*$ indicates the error probability at the equilibrium.}
\label{tab.mixedNASHfixed8m4}
\end{table}

The last case we have analyzed corresponds to a situation in which the FC knows only that the number of Byzantines is lower than $n/2$ (see Sec. \ref{subsec.less_half}). The payoff for this instantiation of the $DF_{Byz}$ game is given in Table \ref{tab.lessN2m4}. In order to compare the results of this case with those obtained for the case of independent nodes and that of fixed number of Byzantines, we observe that when all the sequences $a^n$ with $n_B < n/2$ have the same probability, the average number of Byzantines turns out to be 7.86. The most similar settings, then, are that of independent nodes with $\alpha = 0.4$ and that of fixed number of nodes with $n_B = 8$. With respect to the former, the error probability at the equilibrium is significantly smaller, thus confirming the case of independent nodes as the worst scenario for the FC. This is due to the fact that with $\alpha = 0.4$ it is rather likely that number of Byzantines is larger than 0.5 this making any reliable decision impossible. The error probability obtained with a fixed number of Byzantines equal to 8, however, is much lower. This is a reasonable result, since in that case the a priori information available to the FC permits a better localization of the corrupted reports.

\begin{table}[h!]
\centering
\renewcommand{\arraystretch}{1.1}
\begin{tabular}{c| c| c| c| c| c| c|}
                       \hline
\multicolumn{1}{|c|}{\tiny{$P_{mal}^{B}$/$P_{mal}^{FC}$}} & 0.5   & 0.6   & 0.7   & 0.8   &0.9   &1.0   \\ \hline
\multicolumn{1}{|c|}{0.5}  &0.15 &0.17  &0.20  &0.29 &0.39  &0.51 \\ \hline
\multicolumn{1}{|c|}{0.6}  &0.17 &0.16  &0.16  &0.22 &0.29  &0.40 \\ \hline
\multicolumn{1}{|c|}{0.7}  &0.19 &0.15  &0.14  &0.16 &0.20  &0.30 \\ \hline
\multicolumn{1}{|c|}{0.8}  &0.27 &0.20  &0.17  &0.16 &0.17  &0.22 \\ \hline
\multicolumn{1}{|c|}{0.9}  &0.85 &0.76  &0.72  &0.63 &0.58  &0.63 \\ \hline
\multicolumn{1}{|c|}{1.0}  &3.81 &3.49  &3.30  &2.62 &2.24  & {\bf 2.13} \\ \hline
\end{tabular}

\caption{Payoff of the $DF_{Byz}$ game ($10^2 \times P_e$) with $N_B < n/2$. The  other parameters of the game are set as follows: $m= 4$, $n = 20$, $\varepsilon = 0.1$. The equilibrium point is highlighted in bold.}
\label{tab.lessN2m4}
\end{table}

\subsubsection{Medium values of $m$}

In this section we report the results that we got when the length of the observation vector increases. We expect that by comparing the reports sent by the nodes corresponding to different components of the state vector allows a better identification of the byzantine nodes, thus modifying the equilibrium of the game. Specifically, we repeated the simulations carried out in the previous section, by letting $m = 10$. Though desirable, repeating the simulations with even larger values of $m$ is not possible due to the exponential growth of the complexity of the optimum fusion rule with $m$.

Tables \ref{tab.indip03m10} through \ref{tab.indip045m10} report the payoffs of the game for the case of independent node states. As it can be seen, $P_{mal}^B = 1.0$ is still a dominant strategy for the Byzantines and the profile (1,1) is the unique rationalizable equilibrium of the game. Moreover, the value of $P_e$ at the equilibrium is slightly lower than for $m=4$, when $\alpha=0.3$ and $\alpha=0.4$ (see Tables \ref{tab.indip03m4} and \ref{tab.indip04m4}). Such an advantage disappears when $\alpha=0.45$ (see Table \ref{tab.indip045m4}), since the number of Byzantines is so large that identifying them is difficult even with $m =10$.

\begin{table}[h!]
\centering
\renewcommand{\arraystretch}{1.1}
\begin{tabular}{c| c| c| c| c| c| c|}
                       \hline
\multicolumn{1}{|c|}{\tiny{$P_{mal}^{B}$/$P_{mal}^{FC}$}} & 0.5   & 0.6   & 0.7   & 0.8   &0.9   &1.0   \\ \hline
\multicolumn{1}{|c|}{0.5}  &0.258 &0.28  &0.39  &0.63  &1.0  &1.7 \\ \hline
\multicolumn{1}{|c|}{0.6}  &0.28 &0.226  &0.248  &0.362  &0.652  &2.0 \\ \hline
\multicolumn{1}{|c|}{0.7}  &0.346 &0.22  &0.206  &0.23  &0.314  &5.3 \\ \hline
\multicolumn{1}{|c|}{0.8}  &1.2 &0.648  &0.44  &0.428  &0.498  &13.9 \\ \hline
\multicolumn{1}{|c|}{0.9}  &8.6 &7.8  &7.6  &7.8  &7.5  &19.9 \\ \hline
\multicolumn{1}{|c|}{1.0}  &41.9 &46.7  &50.9  &59.8  &52.2  &\bf{32.9} \\ \hline
\end{tabular}

\caption{Payoff of the $DF_{Byz}$ game ($10^3 \times P_e$) with independent node states with $\alpha = 0.3$, $m= 10$, $n = 20$, $\varepsilon = 0.1$. The equilibrium point is highlighted in bold.}
\label{tab.indip03m10}
\end{table}

\begin{table}[h!]
\centering
\renewcommand{\arraystretch}{1.1}
\begin{tabular}{c| c| c| c| c| c| c|}
                       \hline
\multicolumn{1}{|c|}{\tiny{$P_{mal}^{B}$/$P_{mal}^{FC}$}} & 0.5   & 0.6   & 0.7   & 0.8   &0.9   &1.0   \\ \hline
\multicolumn{1}{|c|}{0.5}  &0.11 &0.13  &0.19  &0.73 &2.16  &0.68 \\ \hline
\multicolumn{1}{|c|}{0.6}  &0.11 &8.32e-2  &9.96e-2  &0.26 &0.67  &1.30 \\ \hline
\multicolumn{1}{|c|}{0.7}  &0.18 &7.66e-2  &6.62e-2  &9.52e-2 &0.18  &4.87 \\ \hline
\multicolumn{1}{|c|}{0.8}  &1.10 &0.60  &0.33  &0.24 &0.28  &10.41 \\ \hline
\multicolumn{1}{|c|}{0.9}  &5.77 &4.75  &3.95  &3.53 &3.41  &13.44 \\ \hline
\multicolumn{1}{|c|}{1.0}  &20.41 &21.26  &22.65  &24.27 &26.21  &\bf{18.72} \\ \hline
\end{tabular}

\caption{Payoff of the $DF_{Byz}$ game ($10^2 \times P_e$) with independent node states with $m= 10$, $n = 20$, $\alpha = 0.4$, $\varepsilon = 0.1$. The equilibrium point is highlighted in bold.}
\label{tab.indip04m10}
\end{table}

\begin{table}[h!]
\centering
\renewcommand{\arraystretch}{1.1}
\begin{tabular}{c| c| c| c| c| c| c|}
                       \hline
\multicolumn{1}{|c|}{\tiny{$P_{mal}^{B}$/$P_{mal}^{FC}$}} & 0.5   & 0.6   & 0.7   & 0.8   &0.9   &1.0   \\ \hline
\multicolumn{1}{|c|}{0.5}  &0.20 &0.23  &0.47  &2.88 &10.92  &1.26 \\ \hline
\multicolumn{1}{|c|}{0.6}  &0.22 &0.18  &0.24  &0.80 &2.85  &2.93 \\ \hline
\multicolumn{1}{|c|}{0.7}  &0.50 &0.19  &0.15  &0.23 &0.65  &10.64 \\ \hline
\multicolumn{1}{|c|}{0.8}  &2.61 &1.24  &0.63  &0.41 &0.59  &20.65 \\ \hline
\multicolumn{1}{|c|}{0.9}  &11.74 &9.28  &7.08  &5.65 &5.21  &25.85 \\ \hline
\multicolumn{1}{|c|}{1.0}  &34.25 &34.94  &36.01  &37.74 &39.87  &\bf{33.17}  \\ \hline
\end{tabular}

\caption{Payoff of the $DF_{Byz}$ game ($10^2 \times P_e$) with independent node states with $\alpha = 0.45$, $m= 10$, $n = 20$, $\varepsilon = 0.1$. The equilibrium point is highlighted in bold.}
\label{tab.indip045m10}
\end{table}

The results of the simulations for the case of fixed number of nodes are given in Tables  \ref{tab.fixed6m10} through \ref{tab.fixed9m10}. With respect to the case of $m = 4$, the optimum strategy for the Byzantines shifts to $P_{mal}^B = 0.5$. When $n_B =6$,  $P_{mal}^B = 0.5$ is a dominant strategy, while for $n_B =8$ and $n_B =9$, no equilibrium point exists if we consider only pure strategies. The mixed strategy equilibrium point for these cases is given in Tables \ref{tab.mixedNASHfixed8m10} and \ref{tab.mixedNASHfixed9m10}. By comparing those tables with those of the case $m=4$, the preference towards $P_{mal}^B = 0.5$ is evident.

\begin{table}[h!]
\centering
\renewcommand{\arraystretch}{1.1}
\begin{tabular}{c| c| c| c| c| c| c|}
                       \hline
\multicolumn{1}{|c|}{\tiny{$P_{mal}^{B}$/$P_{mal}^{FC}$}} & 0.5   & 0.6   & 0.7   & 0.8   &0.9   &1.0   \\ \hline
\multicolumn{1}{|c|}{0.5}  &\bf{1.22} &\bf{1.22}  &1.40  &2.20  &5.06  &11.0 \\ \hline
\multicolumn{1}{|c|}{0.6}  &1.12 &0.94  &1.02  &1.26  &2.56  &5.34 \\ \hline
\multicolumn{1}{|c|}{0.7}  &1.22 &0.58  &0.56  &0.64  &0.98  &2.06 \\ \hline
\multicolumn{1}{|c|}{0.8}  &1.22 &0.36  &0.32  &0.28  &0.30  &0.56 \\ \hline
\multicolumn{1}{|c|}{0.9}  &1.40 &0.20  &0.18  &0.16  &0.10  &0.18 \\ \hline
\multicolumn{1}{|c|}{1.0}  &1.52 &0.14  &0.14  &0.10  &6e-2  &4e-2 \\ \hline
\end{tabular}

\caption{Payoff of the $DF_{Byz}$ game ($10^4 \times P_e$) with $n_B = 6$, $m= 10$, $n = 20$,  $\varepsilon = 0.1$. The equilibrium point is highlighted in bold.}
\label{tab.fixed6m10}
\end{table}

\begin{table}[h!]
\centering
\renewcommand{\arraystretch}{1.1}
\begin{tabular}{c| c| c| c| c| c| c|}
                       \hline
\multicolumn{1}{|c|}{\tiny{$P_{mal}^{B}$/$P_{mal}^{FC}$}} & 0.5   & 0.6   & 0.7   & 0.8   &0.9   &1.0   \\ \hline
\multicolumn{1}{|c|}{0.5}  &4.04 &4.44  &6.24  &10.0 &24.0  &71.0 \\ \hline
\multicolumn{1}{|c|}{0.6}  &4.02 &3.30  &3.58  &5.24 &10.0  &26.0 \\ \hline
\multicolumn{1}{|c|}{0.7}  &3.48 &2.16  &2.14  &2.16 &3.26  &7.76 \\ \hline
\multicolumn{1}{|c|}{0.8}  &3.56 &1.10  &0.88  &0.78 &0.98  &2.08 \\ \hline
\multicolumn{1}{|c|}{0.9}  &4.60 &0.68  &0.54  &0.30 &0.26  &0.44 \\ \hline
\multicolumn{1}{|c|}{1.0}  &5.20 &0.54  &0.20  &8e-2 &0  &0 \\ \hline
\end{tabular}

\caption{Payoff of the $DF_{Byz}$ game ($10^4 \times P_e$) with $n_B = 8$, $m= 10$, $n = 20$,  $\varepsilon = 0.1$. No pure strategy equilibrium exists.}
\label{tab.fixed8m10}
\end{table}

\begin{table}[h!]
\centering
\renewcommand{\arraystretch}{1.1}
\begin{tabular}{c| c| c| c| c| c| c|}
                       \hline
\multicolumn{1}{|c|}{\tiny{$P_{mal}^{B}$/$P_{mal}^{FC}$}} & 0.5   & 0.6   & 0.7   & 0.8   &0.9   &1.0   \\ \hline
\multicolumn{1}{|c|}{0.5}  &6.74 &7.82  &12  &23 &52  &168 \\ \hline
\multicolumn{1}{|c|}{0.6}  &5.44 &4.94  &6.14  &9.40 &18  &52 \\ \hline
\multicolumn{1}{|c|}{0.7}  &4.22 &3.30  &2.78  &3.38 &5.86  &15 \\ \hline
\multicolumn{1}{|c|}{0.8}  &3.0 &2.24  &1.24  &0.78 &1.32  &3.24 \\ \hline
\multicolumn{1}{|c|}{0.9}  &5.22 &2.36  &1.34  &1.02 &0.88  &1.24 \\ \hline
\multicolumn{1}{|c|}{1.0}  &70 &40  &19  &8.90 &3.44  &2.42  \\ \hline
\end{tabular}

\caption{Payoff of the $DF_{Byz}$ game ($10^4 \times P_e$) with $n_B = 9$, $m= 10$, $n = 20$, $\varepsilon = 0.1$. No pure strategy equilibrium exists.}
\label{tab.fixed9m10}
\end{table}

Eventually, Table \ref{tab.lessN2m10}, gives the results for the case $N_B < n/2$. As in the case of fixed number of Byzantines, the equilibrium point strategy passes from the pure strategy (1,1) to a mixed strategy (see Table \ref{tab.mixedNASHlessm10}). Once again, the reason for such a behavior, is that when $m$ increase, the amount of information available to the FC increases, hence making the detection of corrupted reports easier. As a result, the Byzantines must find a trade-off between forcing a wrong decision and reducing the mutual information between the corrupted reports and system states.

We conclude observing that even with $m=10$, the case of independent nodes results in the worst performance.

\begin{table}[h!]
\centering
\renewcommand{\arraystretch}{1.1}
\begin{tabular}{c| c| c| c| c| c| c|}
                       \hline
\multicolumn{1}{|c|}{\tiny{$P_{mal}^{B}$/$P_{mal}^{FC}$}} & 0.5   & 0.6   & 0.7   & 0.8   &0.9   &1.0   \\ \hline
\multicolumn{1}{|c|}{0.5}  &4.46 &5.38  &6.64  &9.88 &16  &27 \\ \hline
\multicolumn{1}{|c|}{0.6}  &3.90 &3.38  &4.10  &5.90 &9.42  &19 \\ \hline
\multicolumn{1}{|c|}{0.7}  &3.04 &2.24  &1.82  &2.26 &3.68  &7.28 \\ \hline
\multicolumn{1}{|c|}{0.8}  &2.78 &1.72  &1.0  &0.72 &0.90  &1.70 \\ \hline
\multicolumn{1}{|c|}{0.9}  &3.24 &1.38  &0.62  &0.30 &0.20  &0.48 \\ \hline
\multicolumn{1}{|c|}{1.0}  &27 &15  &6.84  &4.68 &1.42  &1.04 \\ \hline
\end{tabular}

\caption{Payoff of the $DF_{Byz}$ game ($10^4 \times P_e$) with $N_B < n/2$. The  other parameters of the game are set as follows: $m= 10$, $n = 20$, $\varepsilon = 0.1$. No pure strategy equilibrium exists.}
\label{tab.lessN2m10}
\end{table}

\begin{table}[h!]
\centering
\renewcommand{\arraystretch}{1.4}
\begin{tabular}{c| c| c| c| c| c| c|}
                       \hline
\multicolumn{1}{|c|}{}& 0.5   & 0.6   & 0.7   & 0.8   &0.9   &1.0   \\ \hline
\multicolumn{1}{|c|}{$P(P_{mal}^{B})$ }  & 0.921 & 0 & 0 & 0 & 0 & 0.079 \\ \hline
\multicolumn{1}{|c|}{$P(P_{mal}^{FC})$ }  &0.771 &0.229  & 0  &0 & 0 &0 \\ \hline
\multicolumn{7}{|c|}{$P_e^* = 4.13e-4$} \\ \hline
\end{tabular}

\caption{Mixed equilibrium point for the $DF_{Byz}$ game with  $n_B = 8$, $m= 10$, $n = 20$,  $\varepsilon = 0.1$. $P_e^*$ indicates the error probability at the equilibrium.}
\label{tab.mixedNASHfixed8m10}
\end{table}

\begin{table}[h!]
\centering
\renewcommand{\arraystretch}{1.4}
\begin{tabular}{c| c| c| c| c| c| c|}
                       \hline
\multicolumn{1}{|c|}{}& 0.5   & 0.6   & 0.7   & 0.8   &0.9   &1.0   \\ \hline
\multicolumn{1}{|c|}{$P(P_{mal}^{B})$ }  & 0.4995 & 0 & 0 & 0 & 0 & 0.5005 \\ \hline
\multicolumn{1}{|c|}{$P(P_{mal}^{FC})$ }  &0 &0  &0.66  &0.34 & 0 &0 \\ \hline
\multicolumn{7}{|c|}{$P_e^* = 1.58e-3$} \\ \hline\end{tabular}

\caption{Mixed equilibrium point for the $DF_{Byz}$ game with $n_B = 9$, $m= 10$, $n = 20$,  $\varepsilon = 0.1$. $P_e^*$ indicates the error probability at the equilibrium.}
\label{tab.mixedNASHfixed9m10}
\end{table}

\begin{table}[h!]
\centering
\renewcommand{\arraystretch}{1.4}
\begin{tabular}{c| c| c| c| c| c| c|}
                       \hline
\multicolumn{1}{|c|}{}& 0.5   & 0.6   & 0.7   & 0.8   &0.9   &1.0   \\ \hline
\multicolumn{1}{|c|}{$P(P_{mal}^{B})$ }  & 0.4 & 0 & 0 & 0 & 0 & 0.6 \\ \hline
\multicolumn{1}{|c|}{$P(P_{mal}^{FC})$ }  &0 &0  &0.96  &0.04 & 0 &0 \\ \hline
\multicolumn{7}{|c|}{$P_e^* = 6.76e-4$} \\ \hline
\end{tabular}

\caption{Mixed equilibrium point for the $DF_{Byz}$ game with $N_B < n/2$ with $m= 10$, $n = 20$, $\varepsilon = 0.1$. $P_e^*$ indicates the error probability at the equilibrium.}
\label{tab.mixedNASHlessm10}
\end{table}

\subsection{Performance at the equilibrium and comparison with prior works}

As a last analysis we compare the error probability obtained by game-theoretic optimum decision fusion introduced in this paper, with those obtained by previous solutions. Specifically, we compare our scheme against a simple majority-based decision fusion rule according to which the FC decides that  $s_j = 1$ if and only if $\sum_i r_{ij} > n/2$ ($\mathsf{Maj}$), against the hard isolation scheme described in \cite{Raw11} ($\mathsf{HardIS}$), and the soft isolation scheme proposed in \cite{CDC} ($\mathsf{SoftIS}$).

In order to carry out a fair comparison and to take into account the game-theoretic nature of the problem, the performance of all the schemes are evaluated at the equilibrium. For the $\mathsf{HardIS}$ and $\mathsf{SoftIS}$ schemes this corresponds to letting $P_{mal}^{B}$ = 1. In fact, in \cite{CDC}, it is shown that this is a dominant strategy for these two specific fusion schemes. As a consequence, $P_{mal}^{FC}$ is also set to 1, since the FC knows in advance that the Byzantines will play the dominant strategy. For the $\mathsf{Maj}$ fusion strategy, the FC has no degrees of freedom, so no game actually exists in this case. With regard to the Byzantines, it is easy to realize that the best strategy is to let $P_{mal}^B = 1$. When the equilibrium corresponds to a mixed strategy, the error probability is averaged according to the mixed strategies at the equilibrium. Tables \ref{tab.payoffCOMPm4} and \ref{tab.payoffCOMPm10} show the error probability at the equilibrium for the tested systems under different setups. As it can be seen, the fusion scheme resulting for the application of the optimum fusion rule in a game-theoretic setting, consistently provides better results for all the analyzed cases. Expectedly, the improvement is more significant for the setups in which the FC has more information about the distribution of the Byzantines across the network.

\begin{table}[h!]
\centering
\renewcommand{\arraystretch}{1.1}
\begin{tabular}{c| c| c| c| c|}
                       \hline
\multicolumn{1}{|l|}{} & $\mathsf{Maj}$   & $\mathsf{HardIS}$   & $\mathsf{SoftIS}$   & $\mathsf{OPT}$   \\ \hline
\multicolumn{1}{|l|}{Independent nodes, $\alpha = 0.3$}  & 0.073 & 0.048 & 0.041 & 0.035 \\ \hline
\multicolumn{1}{|l|}{Independent nodes, $\alpha = 0.4$}  & 0.239 &  0.211 & 0.201 & 0.192\\ \hline
\multicolumn{1}{|l|}{Independent nodes, $\alpha = 0.45$}   & 0.362& 0.344 & 0.338 & 0.331\\ \hline
\multicolumn{1}{|l|}{Fixed n. of nodes $n_B = 6$}  & 0.017 & 0.002 & 6.2e-4 & 3.8e-4\\ \hline
\multicolumn{1}{|l|}{Fixed n. of nodes $n_B = 8$}   & 0.125 & 0.044 & 0.016 & 0.004\\ \hline
\multicolumn{1}{|l|}{Fixed n. of nodes $n_B = 9$}  & 0.279 & 0.186  & 0.125 & 0.055\\ \hline
\multicolumn{1}{|l|}{Max entropy with $N_B  < n/2$}  & 0.154 & 0.086 & 0.052 & 0.021\\ \hline
\end{tabular}

\caption{Error probability at the equilibrium for various fusion schemes. All the results have been obtained by letting $m=4$, $n = 20$, $\varepsilon = 0.1$.}
\label{tab.payoffCOMPm4}
\end{table}

\begin{table}[h!]
\centering
\renewcommand{\arraystretch}{1.1}
\begin{tabular}{c| c| c| c| c|}
                       \hline
\multicolumn{1}{|l|}{} & $\mathsf{Maj}$   & $\mathsf{HardIS}$   & $\mathsf{SoftIS}$   & $\mathsf{OPT}$   \\ \hline
\multicolumn{1}{|l|}{Independent nodes, $\alpha = 0.3$}  &0.073 &0.0364  &0.0346  &0.033  \\ \hline
\multicolumn{1}{|l|}{Independent nodes, $\alpha = 0.4$}  &0.239 &0.193  &0.19  &0.187 \\ \hline
\multicolumn{1}{|l|}{Independent nodes, $\alpha = 0.45$}   &0.363 &0.334  &0.333  &0.331 \\ \hline
\multicolumn{1}{|l|}{Fixed n. of nodes $n_B = 6$}  &0.016 &1.53e-4  &1.41e-4  &1.22e-4 \\ \hline
\multicolumn{1}{|l|}{Fixed n. of nodes $n_B = 8$}   &0.126 &0.0028  &9.68e-4  &4.13e-4 \\ \hline
\multicolumn{1}{|l|}{Fixed n. of nodes $n_B = 9$}  &0.279 &0.0703  &0.0372  &1.58e-3 \\ \hline
\multicolumn{1}{|l|}{Max entropy with $N_B  < n/2$}  &0.154 &0.0271  &0.0141  &6.8e-4 \\ \hline
\end{tabular}

\caption{Error probability at the equilibrium for various fusion schemes. All the results have been obtained by letting $m=10$, $n = 20$, $\varepsilon = 0.1$.}
\label{tab.payoffCOMPm10}
\end{table}

\section{Conclusions}
\label{sec.conc}

We have studied the problem of decision fusion in multi-sensor networks in the presence of Byzantines. We first derived the optimum decision strategy by assuming that the statistical behavior of the Byzantines is known. Then we relaxed such an assumption by casting the problem into a game-theoretic framework in which the FC tries to guess the behavior of the Byzantines. The Byzantines, in turn, must fix their corruption strategy without knowing the guess made by the FC. We considered several versions of the game with different distributions of the Byzantines across the network. Specifically, we considered three setups: unconstrained maximum entropy distribution, constrained maximum entropy distribution and fixed number of Byzantines. In order to reduce the computational complexity of the optimum fusion rule for large network sizes, we proposed an efficient implementation based on dynamic programming. Simulation results show that increasing the observation window $m$ leads to better identification of the Byzantines at the FC. This forces the Byzantines to look for a  trade-off between forcing the FC to make a wrong decision on one hand, and reducing the mutual information between the reports and the system state on the other hand. Simulation results confirm that, in all the analyzed cases, the performance at the equilibrium are superior to those obtained by previously proposed techniques. 
About the future work, an interesting possibility is to enhance the Byzantines performance by granting them access to the observation vectors. In this way, 
they can focus their attack on the most uncertain cases thus avoiding to flip the local decision when it is expected that the attack will have no effect on the FC decision.
Considering a case where the node can send more extensive reports rather than one single bit is another interesting possibility. 
Finally, a motivating direction for future research is to investigate the adversarial decision fusion in a distributed scenario.

\bibliographystyle{IEEEtran}
\bibliography{DFusionByz}

\end{document}